\documentclass[preprint,3p]{elsarticle}

\usepackage{amssymb}
\usepackage[version=3]{mhchem}
\usepackage{hyperref}
\usepackage{graphicx}
\usepackage{subcaption}
\usepackage{threeparttablex}
\usepackage{tabularx}
\usepackage{bm}
\usepackage{multirow}
\usepackage{makecell}

\journal{Nuclear Engineering and Technology}

\begin{document}

\begin{frontmatter}



\title{Waveform-based pulse timing enhanced by self-supervised neural regression models for digital nuclear instrumentation}


\author[myfirstaff,mysecondaff]{Pengcheng Ai\corref{mycorrespondingauthor}}
\ead{aipc@ccnu.edu.cn}

\author[myfirstaff,mysecondaff]{Xiangming Sun}

\cortext[mycorrespondingauthor]{Corresponding author}

\affiliation[myfirstaff]{organization={PLAC, Key Laboratory of Quark and Lepton Physics (MOE), Central China Normal University},
            addressline={No. 152 Luoyu Road}, 
            city={Wuhan},
            postcode={430079}, 
            state={Hubei},
            country={China}}

\affiliation[mysecondaff]{organization={Hubei Provincial Engineering Research Center of Silicon Pixel Chip \& Detection Technology},
	addressline={No. 152 Luoyu Road}, 
	city={Wuhan},
	postcode={430079}, 
	state={Hubei},
	country={China}}

\begin{abstract}
With the rapid development of high-sampling-rate digital nuclear devices, their applications in nuclear facilities and related fields become more acceptable, and will possibly become widespread due to the versatility and flexibility of waveform sampling systems. When digitizers reach hundreds of megahertz or gigahertz, their lower limit in time measurement probes into the sub-nanosecond region, expediting their use in a bunch of time-critical scenarios. While conventional algorithms coupled with digital sampling techniques realize fair pulse timing in certain conditions, the potentials are still open to compact and efficient machine learning models. In this paper, we intend to boost the performance of pulse timing with a new paradigm of self-supervised learning on matched waveform pairs. The proposed method combines the mainstream self-supervised loss and the auxiliary weakly-supervised loss on the reference set, jointly improving the timing performance. We apply the method in a gamma/neutron-mixed radiation field, common in nuclear facilities. The simulation and experimental results show that, the proposed method not only outperforms conventional timing algorithms, but also surpasses neural network models based on supervised learning alone. Considering the trained model is lightweight and computationally efficient, it is promising to deploy it on edge devices, like economical microcontrollers with neural processing units.
\end{abstract}


\end{frontmatter}


\section{Introduction}

In nuclear facilities, measuring the arrival time of incident particles can provide important information about the monitored reactions and assist real-time multi-dimensional observation. For example, at Chinese Spallation Neutron Source (CSNS), the time-of-flight (ToF) of back-streaming white neutrons is measured to obtain the energy spectrum for neutron-related experiments \cite{8651570}. The detector is calibrated with a waveform sampling system at 1 GSps \cite{8-20201823} and measured by switched capacitor arrays to reach time resolution of $\sim$10 ns \cite{202006-5406-yzn08}. The digital data acquisition (DAQ) system for Versatile Array of Neutron Detectors at Low-Energy (VANDLE) includes digitizers with the sampling period at the nanosecond scale and reaches timing resolution below nanoseconds by pulse shape analysis \cite{PAULAUSKAS201422}. Besides, the time measurement enables discrimination between neutrons and gammas in a mixed radiation field \cite{LI201330,alharbi2024wavelet,YAOXUAN2025112091}, which is critical in fusion devices (e.g. tokamaks), accelerator‑driven sub-critical systems, and radiation monitoring systems for accelerator‑based neutron sources. Furthermore, in nuclear safeguards, fast timing correlation facilitates three-dimensional imaging of special nuclear materials (SNM) \cite{Jana2021}, accurate assays of fuel assemblies in nuclear fuel fabrication plants \cite{BEAUMONT2020163682}, and radiation portal monitoring for SNM \cite{murer2013flash}. When digitizers reach $\sim$100 Msps to $\sim$GSps, the DAQ systems in nuclear facilities will benefit substantially to realize sub-ns timing resolution, with the potential to reform the principle and techniques of measurement in these applications.

With the high-speed digitizers discussed above, pulse timing can be applied in the digital domain, with general-purpose devices or application-specific integrated circuits. Conventional timing algorithms \cite{NELSON2003324,4337024,FAISAL2011479} include leading-edge discrimination (LED), digital constant fraction discrimination (CFD), pulse shape analysis based on peak, maximum slope or linear extrapolation, and pulse matching methods such as cross correlation. On the other hand, with the development of artificial intelligence, especially data-driven machine learning techniques, the capability of digital pulse timing has been unleashed to approach the theoretical limit \cite{Ai_2021}. The performance of optimization-based neural networks is foreseeable and guaranteed with little human intervention.

However, the precision of supervised learning models relies heavily on the quality of timing labels, which is usually hard to be accurately determined without sophisticated timing devices (with the resolution at least one order of magnitude higher than required). In this paper, we propose a new training paradigm to enhance the time resolution and to ease the demands for timing labels by specially designed self supervision with matched waveform pairs. The contributions of the paper include:

\begin{itemize}
	\item A new loss function for digital pulse timing, combining self-supervised learning on matched waveform pairs and weakly-supervised learning on a few labelled reference examples.
	\item A thorough simulation study, from conceptual proof to experimental approximation, demonstrating the desired performance of the proposed loss function.
	\item Quantitative validation of the method on experimental data without actual timing labels, showing the outstanding timing resolution in a circumstance of asynchronous detection.
\end{itemize}

\subsection{Related work}

In medical imaging, convolutional neural networks (CNN) were used for ToF estimation on positron emission tomography (PET) detectors \cite{Berg_2018}. Then, pulse timing with neural networks in nuclear radiation detectors was proposed with a comparison to curve fitting \cite{Ai_2019}. Ultra fast timing with neural networks made PET without explicit reconstruction possible \cite{Kwon_2021}. CNN was also utilized to correct the timing errors of LED to reduce its bias in timing \cite{Onishi_2022}. CNN was further employed for timing analysis of the photomultiplier tube (PMT) waveform on ToF detectors \cite{10038575}. Customized digital logic of neural network inference for pulse timing was achieved by template-based hardware architectures \cite{AI2020164420,10005128}. Finally, the self-supervised learning method was devised on modularized nuclear detectors for pulse timing free from label annotations \cite{Ai_2023}.

Beyond pulse timing, neural networks have been widely applied in nuclear instrumentation for energy resolution optimization \cite{10666769,HE2026104337}, pulse shape parameterization \cite{WANG2023110787,Ai_2025}, neutron/gamma discrimination \cite{GUO2025107400,AI2025103667}, pile-up recognition \cite{PAN2025103329,10190588}, etc.

\section{Method}

\subsection{An overview of self-supervised pulse timing}

We begin by briefly reviewing the theoretical results in Ref. \cite{Ai_2023}. Suppose we have a multi-channel detector with linear correlations in arrival time. For each channel, we sample a series of waveform points:
\begin{equation}
	\bm{s}_i = \left( s_i^{(0)}, s_i^{(1)}, \dots, s_i^{(K-1)} \right),\ \ i = 0, 1, \dots, N-1
\end{equation}

\noindent where $K$ is the length of the waveform points, and $N$ is the number of channels. We feed the time series in each channel into a neural network and get a vector of predicted arrival time:
\begin{equation}
	y_i = f_{\bm{\eta}}(\bm{s}_i) - \Delta t_i
\end{equation}
\begin{equation}
	\bm{y} = \left( y_0, y_1, \dots, y_{N-1} \right)^\top
\end{equation}

\noindent where $f_{\bm{\eta}}(\cdot)$ is the mapping function of the neural network with weights $\bm{\eta}$, and $\Delta t_i$ is an additional time shift applied to the waveform points, working as a regularizer to prevent the neural network converging to the constant function. It has been proven that, if we minimize the loss expression below, the neural network will learn a function linear to the arrival time:
\begin{equation}
	I(\bm{y}) = \left| \left| \left( \bm{I} - \bm{A}(\bm{A}^\top \bm{A})^{-1} \bm{A}^\top \right) \bm{y} \right| \right|_2^2 = \bm{y}^\top \bm{M} \bm{y} \label{equ:self-sup-loss}
\end{equation}
\begin{equation}
	\text{where:}\quad \bm{A} = \begin{bmatrix}
		1 & 1 & \cdots & 1 \\
		a_0 & a_1 & \cdots & a_{N-1}
	\end{bmatrix}^\top,\quad \bm{M} = \bm{I} - \bm{A}(\bm{A}^\top \bm{A})^{-1} \bm{A}^\top
\end{equation}

\noindent and $(a_0,\ a_1,\ \dots, a_{N-1})$ are geometrical coefficients. The final mapping function of the neural network will be:
\begin{equation}
	f_{\bm{\eta}}(\bm{s}_{*}) = f_{*}(t_{*}) = kt_{*} + b \label{equ:self-sup-time}
\end{equation}

\noindent where $f_{*}(\cdot)$ is a function of the ground-truth arrival time ($t_{*}$) associated with $\bm{s}_{*}$, $k$ is a positive constant less than one, and $b$ is an arbitrary constant.

The above principle of self-supervised learning works well with the multi-channel detector with linear timing correlations, because the timing uncertainties are independent in each channel. However, if we apply the same method to a single-channel detector, there are some noticeable differences. We will discuss why the self-supervised learning method cannot be directly applied to the single-channel case.

If the detector has only one channel, we can get two series of waveform points with a random time shift in between:
\begin{gather}
	\bm{s}_0 = \left( s_0^{(0)}, s_0^{(1)}, \dots, s_0^{(K-1)} \right) \label{equ:wave-pair-0} \\
	\bm{s}_1 = \left( s_1^{(0)}, s_1^{(1)}, \dots, s_1^{(K-1)} \right) \label{equ:wave-pair-1} \\
	\Delta t = (a_1 - a_0) t_c \label{equ:wave-pair-d}
\end{gather}

\noindent where $(a_0,\ a_1)$ are starting points of the two series, and $t_c$ is the sampling period. In equation (\ref{equ:self-sup-loss}), if we set $\bm{M}$ to represent the timing correlations:
\begin{equation}
	\bm{M} = \begin{bmatrix}
		1 & -1 \\
		-1 & 1
	\end{bmatrix}
\end{equation}

Then, the loss expression will be:
\begin{equation}
	I(\bm{y}) = \bm{y}^\top \bm{M} \bm{y} = (y_0 - y_1)^2 = (f_{\bm{\eta}}(\bm{s}_0) - f_{\bm{\eta}}(\bm{s}_1) - \Delta t)^2 \label{equ:self-sup-two}
\end{equation}

If we still substitute equation (\ref{equ:self-sup-time}) into equation (\ref{equ:self-sup-two}), we can find that the term $b$ is cancelled:
\begin{equation}
	I(\bm{y}) = \left( (kt_0 + b) - (kt_1 + b) - \Delta t \right)^2 = \left( k(t_0 - t_1) - \Delta t \right)^2 
\end{equation}

In other words, the loss expression does not put a restriction on the constant term $b$. We call it the \emph{time base drift} problem. The problem will cause the discrepancies between the time base of different waveform examples. Although for each example, we can get a desired linear function, the examples in the dataset do not have a unified time base, which will hinder subsequent timing analysis. Please refer to Section~\ref{sec:poc} for a detailed demonstration.

\subsection{Main method}

\begin{figure}[htb]
	\centering
	\includegraphics[width=0.98\textwidth]{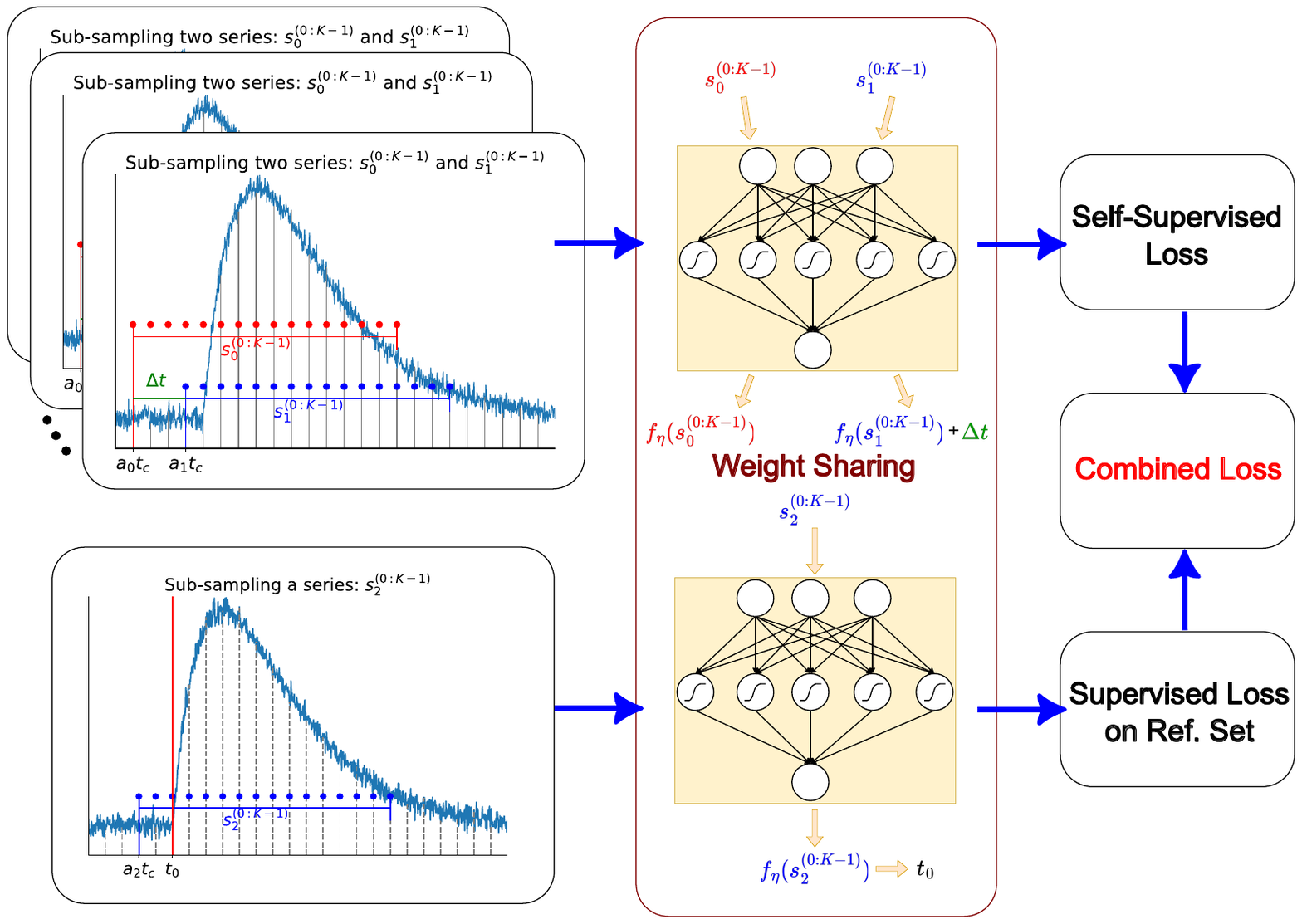}
	\caption{\label{fig:method} The proposed training paradigm combining self-supervised learning on matched waveform pairs and weakly-supervised learning on a few waveform examples for reference.}
\end{figure}

To overcome the time base drift, we propose a simple but effective method combining self-supervised learning and weakly-supervised learning. The term ``weakly-supervised'' arises from the fact that only a small portion of the dataset is annotated with timing labels when the training paradigm is adopted.

The functional diagram of the proposed method is shown in Fig.~\ref{fig:method}. First, on the training set of dataset, self-supervised learning is performed with respect to each example. A waveform pair with different starting points is subsampled from the original waveform points in this example. We define a range at the beginning part of the waveform as possible starting points, and random shifts are generated for each one in the pair:
\begin{gather}
	\hat{a} \sim \mathcal{U}(a_l, a_h) \\
	x_0,\ x_1 \sim \mathcal{B}(0.5) \\
	a_0 = \hat{a} + (2x_0 - 1) \cdot \tilde{a} \\
	a_1 = \hat{a} + (2x_1 - 1) \cdot \tilde{a}
\end{gather}

\noindent where $\hat{a}$ is the common mode of the starting points sampled in the range $(a_l,\ a_h)$ from a uniform distribution, $(x_0,\ x_1)$ are random shifts sampled from the Bernoulli distribution, and $\tilde{a}$ is the amount (usually set to 1) in sampling periods for random shifts. The subsampling process produces a matched waveform pair in equation (\ref{equ:wave-pair-0}) and equation (\ref{equ:wave-pair-1}), and a time distance in equation (\ref{equ:wave-pair-d}).

For self-supervised learning, the loss function would be:
\begin{equation}
	I_1 (\bm{\eta}) = \frac{1}{M_1} \sum\limits_{j=0}^{M_1-1} (f_{\bm{\eta}}(\bm{s}_{j,0}) - f_{\bm{\eta}}(\bm{s}_{j,1}) - \Delta t_j)^2 \label{equ:self-sup-loss-func}
\end{equation}

\noindent where $(\bm{s}_{j,0},\ \bm{s}_{j,1})$ and $\Delta t_j$ are the subsampled matched waveform pair and the time distance from the j-th example, and $M_1$ is the number of examples in the training set.

For the part of weakly-supervised learning, a reference set is constructed by waveform examples and associated timing labels. The cost to construct the reference set is kept to a minimum, since the main purpose of the reference set is to provide a unified time base instead of working as the optimization target. The subsampling process is also performed on the reference set, and the timing labels are rebased by their distances to the starting points:
\begin{gather}
	a_2 \sim \mathcal{U}(a_l, a_h) \\
	t_2 = t_0 - a_2 t_c
\end{gather}

\noindent where $a_2$ is the starting point sampled in the range $(a_l,\ a_h)$ from a uniform distribution, and $t_2$ is the rebased timing label. The corresponding loss function would be:
\begin{equation}
	I_2 (\bm{\eta}) = \frac{1}{M_2} \sum\limits_{j=0}^{M_2-1} (f_{\bm{\eta}}(\bm{s}_{j,2}) - t_{j,2})^2
\end{equation}

\noindent where $\bm{s}_{j,2}$ and $t_{j,2}$ are the subsampled waveform and the rebased timing label from the j-th example, and $M_2$ is the number of examples in the reference set.

Finally, the loss function combining self-supervised learning and weakly-supervised learning would be:
\begin{equation}
	I (\bm{\eta}) = I_1 (\bm{\eta}) + \lambda \cdot I_2 (\bm{\eta}) \label{equ:hybrid-loss-func}
\end{equation}

\noindent where $\lambda$ is a coefficient controlling the intensity of the reference loss function. It should be noted that, the self-supervised loss and the weakly-supervised loss share the neural network architecture and its weights, jointly optimizing the network towards the desired mapping function.

In practice, the size of the reference set can be much smaller than the size of the training set (i.e. $M_2 \ll M_1$). While the timing precision is determined both by the self-supervised part and the weakly-supervised part, the self-supervised loss is the main optimization target and has higher importance.

\section{Simulation results}

\begin{figure}[htb]
	\centering
	\includegraphics[width=0.98\textwidth]{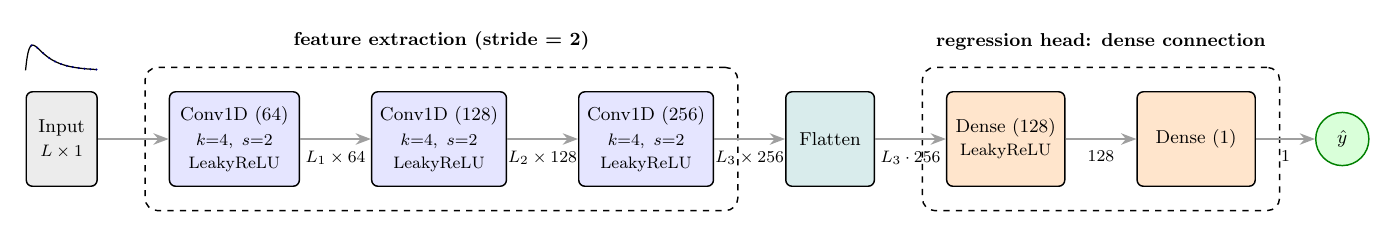}
	\caption{\label{fig:cnn-arch} The architecture of the neural network used throughout the simulation and the experiment. The input time series is fed to three successive one-dimensional convolution layers. Each convolution layer has a number of kernels (64, 128, 256) with size of 4 and stride of 2, followed by Leaky ReLU \cite{pmlr-v238-guo24c} activation function. Afterwards, the feature map is flattened, and two densely connected layers with size of 128 and 1 transform the feature vector to a scalar.}
\end{figure}

Throughout the simulation and the experiment, we choose a CNN architecture as the regression model, shown in Fig.~\ref{fig:cnn-arch}. We use TensorFlow \cite{DBLP:conf/osdi/AbadiBCCDDDGIIK16} along with Keras \cite{chollet2015keras} as our deep learning framework. The Adam \cite{DBLP:journals/corr/KingmaB14} optimizer is chosen to optimize the weights in CNN. The batch size for training is 64. The network training and inference are deployed on a graphics processing unit (NVIDIA RTX 4060 Ti) with video memory of 16 GB.

\subsection{Proof-of-concept simulation}
\label{sec:poc}

\begin{figure}[htb]
	\centering
	\includegraphics[width=0.98\textwidth]{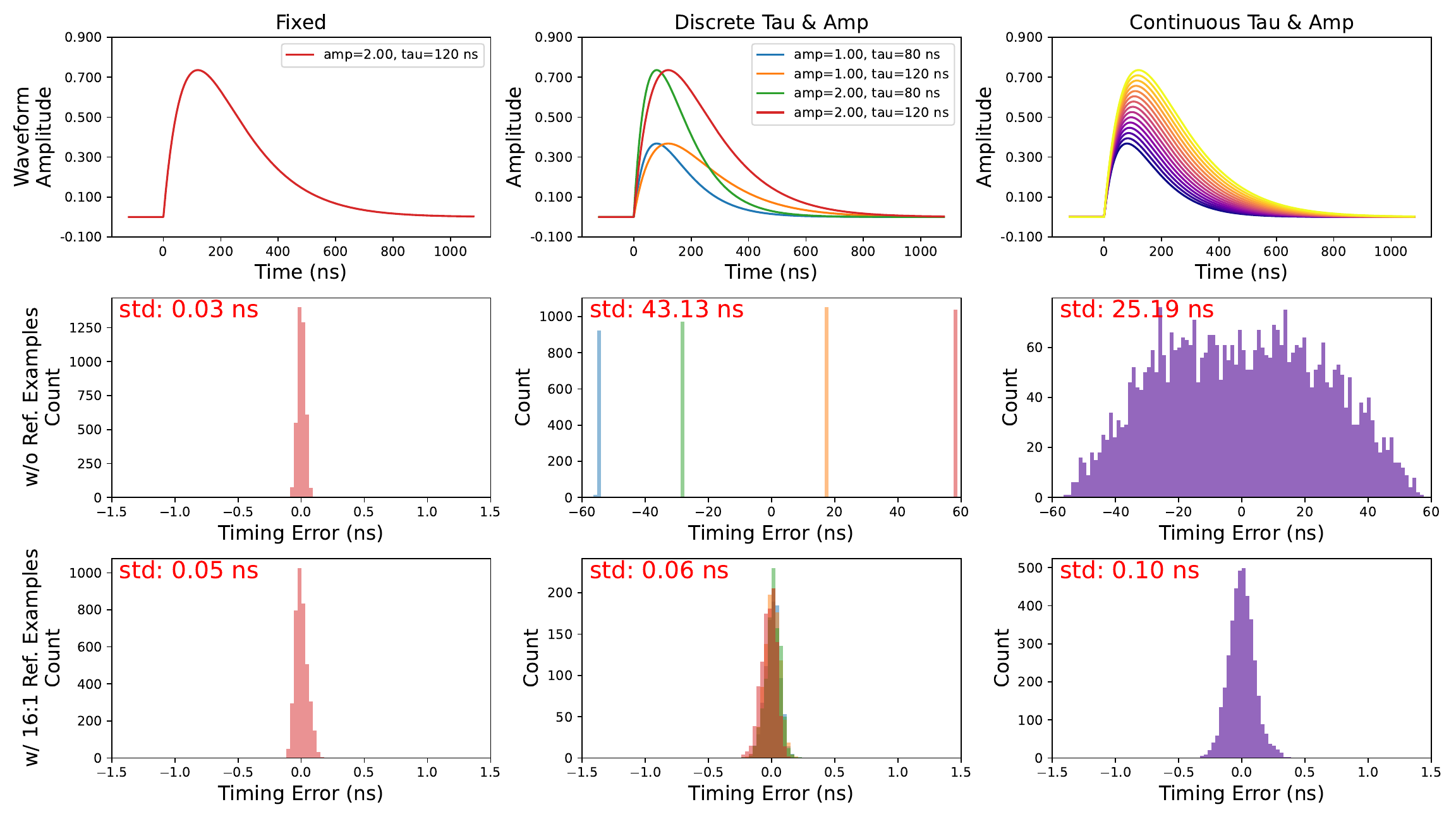}
	\caption{\label{fig:poc-sim} The timing results of the proof-of-concept simulation. Left column: single set of the pulse parameter; mid column: discrete sets of the pulse parameters; Right column: continuous sets of the pulse parameters.}
\end{figure}

A proof-of-concept simulation is conducted to demonstrate the necessity and proper functioning of the hybrid learning method. The nuclear pulses filtered after the capacitance-resistance resistance-capacitance (CRRC) shaper take the following analytical form:
\begin{equation}
	s(t) = A \left( \frac{t-t_0}{\tau} \right) e^{- (t - t_0)/\tau} u(t - t_0)
\end{equation}

\noindent where $t_0$ is the arrival time being predicted, and $(A,\ \tau)$ are waveform parameters. Three different conditions are simulated:

\begin{itemize}
	\item[1.] The waveform parameters $(A,\ \tau)$ take a single fixed set of values.
	\item[2.] The waveform parameters $(A,\ \tau)$ take several discrete sets of values.
	\item[3.] The waveform parameters $(A,\ \tau)$ take continuous sets of values.
\end{itemize}

For each condition, we generate 20000 examples, in which 16000 examples (with 1000 reference examples) are used for training and 4000 examples are used for testing. The waveform is sampled at 1 GSps. The length of the input time series is 1024. A random range of 20 ns is applied to the starting point of each waveform. White Gaussian noise with 0.001 standard deviation is added to waveform points to account for measurement uncertainties. The loss function has been substantially reduced before we finish training.

The simulation results are shown in Fig.~\ref{fig:poc-sim}. The left column, mid column, and right column correspond to three different conditions. In the top row, we display the parameterized waveform in the time domain. In the mid row and bottom row, we compare the test results between using the self-supervised learning alone (equation (\ref{equ:self-sup-loss-func})) and using self-supervised learning along with weakly-supervised learning (equation (\ref{equ:hybrid-loss-func})). For weakly-supervised learning, we use 1 reference example per 16 training examples to regulate the self supervision.

It can be seen in this figure that, the necessity of using weakly-supervised learning on reference examples comes from variations of waveform forms. When the waveform parameters are fixed to a single set of values, self-supervised learning alone can already reach ideal timing resolution. This is because, if the waveform has a fixed form, the time base is common to all examples, and the neural network has the least requirements to adapt to different waveform parameters. As a result, the time base drift does not manifest as a problem in this condition.

In contract, when the waveform parameters take several discrete sets of values, self-supervised learning alone incurs serious time base drift, as shown in the central sub-figure. The distributional centers for 4 different sets of values locate far away from each other. Although each distribution is relatively concentrated, the overall timing performance is greatly compromised. The same issue applies to the condition when waveform parameters take continuous sets of values. In the mid-right sub-figure, we can see the timing errors scatter on a wide range. These conditions demonstrate that the problem of time base drift does exist when the neural network is unable to unite the time base of examples with different waveform parameters.

On the other hand, it can be seen that, the problem of time base drift has been significantly mitigated with the help of weak supervision. With only 16:1 reference examples, the distributional centers in condition 2 have been forced to overlap, and the timing resolution has been greatly improved (from 43.13 ns to 0.06 ns). For condition 3, the improvement of timing resolution is also considerable (from 25.19 ns to 0.10 ns). The phenomena in the proof-of-concept simulation demonstrate in total that, when variations of the waveform forms are apparent, using self-supervised learning along with weakly-supervised learning can compensate for the waveform variations and render the regression model a unified time base.

\subsection{Simulation with the \emph{scintiPulses} package}

\begin{table}[htb]
	\centering
	\caption{\label{tab:scint-fit} Mean parameters of the \emph{scintiPulses} model fitted from experimental data. $\tau_1$ and $\tau_2$ are fast and slow components of the decay model, $p_2$ is the ratio of the slow component, and $\tau_S$ is the standard deviation of the causal Gaussian filter used to simulate the response of a PMT.}
	\begin{tabularx}{0.9\textwidth}{>{\raggedright\arraybackslash}X>{\centering\arraybackslash}X>{\centering\arraybackslash}X>{\centering\arraybackslash}X>{\centering\arraybackslash}X}
		\hline
		Category & $\tau_1$ (ns) & $\tau_2$ (ns) & $p_2$ & $\tau_S$ (ns) \\
		\hline
		Gamma & 8.01 & 103.7 & 0.076 & 2.32 \\
		Neutron & 8.24 & 89.9 & 0.456 & 2.36 \\
		\hline
	\end{tabularx}
\end{table}

\begin{figure}[htb]
	\centering
	\includegraphics[width=0.98\textwidth]{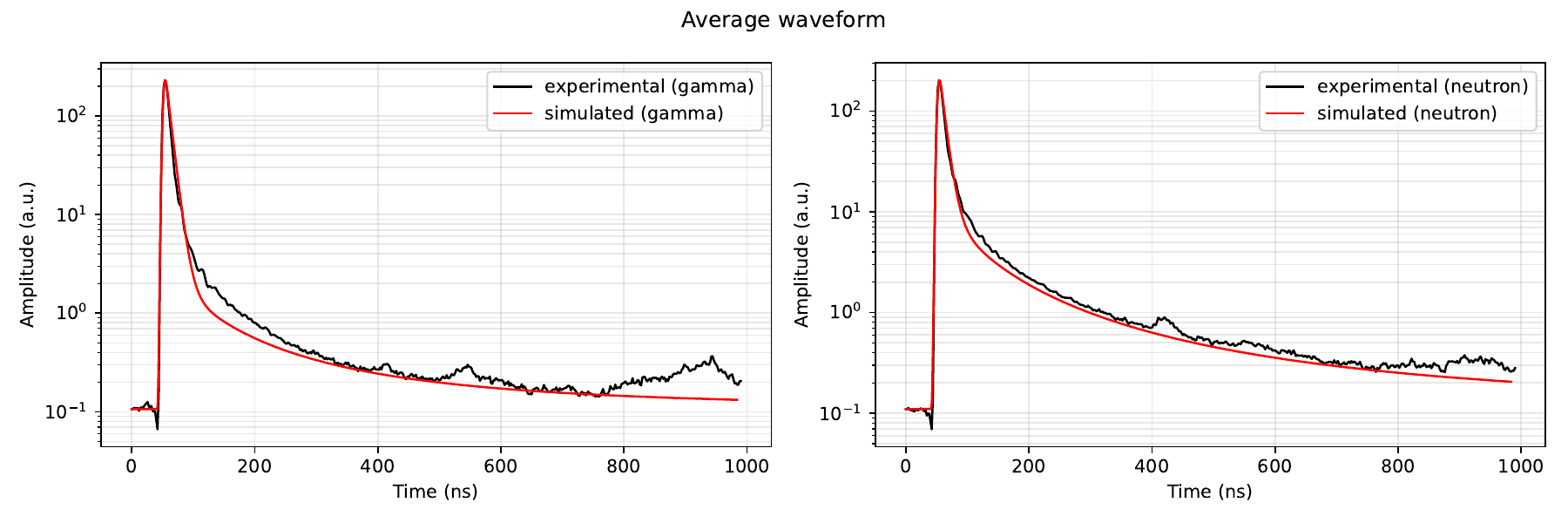}
	\caption{\label{fig:avg-wave} A comparison on the average waveforms from the \emph{scintiPulses} simulation and the experiment.}
\end{figure}

In this section, we perform a simulation study with the \emph{scintiPulses} Python package \cite{Coulon2025}. \emph{scintiPulses} can provide realistic scintillation pulses and is able to produce intermediate outputs from idealized scintillation signals to digitized signals after analog-to-digital converters. Since we aim to generate signals with shape and distribution similar to experimental data (Section~\ref{sec:exp}), a semi-deterministic model generating the PMT anode signals is built.

We consider a mixed radiation field where both gamma events and neutron events are present. In Table~\ref{tab:scint-fit}, we list the mean parameters of the \emph{scintiPulses} model fitted from experimental data. Although there are some literatures discussing the empirical parameters of the scintillator \cite{Iwanowska-Hanke_2014}, we still intend to fit them from scratch, because the similarity to experimental data is prioritized over empirically ``correct'' parameters. In Fig.~\ref{fig:avg-wave}, it can be seen that the simulated waveform approaches the experimental waveform closely. The normalized root mean square errors (NRMSE) for gamma events at the peak region and the tail region are 10.75\% and 7.90\%; for neutron events, the NRMSEs are 9.49\% and 8.55\%. To account for the variations of waveform parameters, we use a Gaussian mixture model to generate the data distribution of parameters, and utilize the \emph{scintiPulses} model to generate waveform examples in the simulation dataset.

\begin{figure}[htb]
	\centering
	\includegraphics[width=0.98\textwidth]{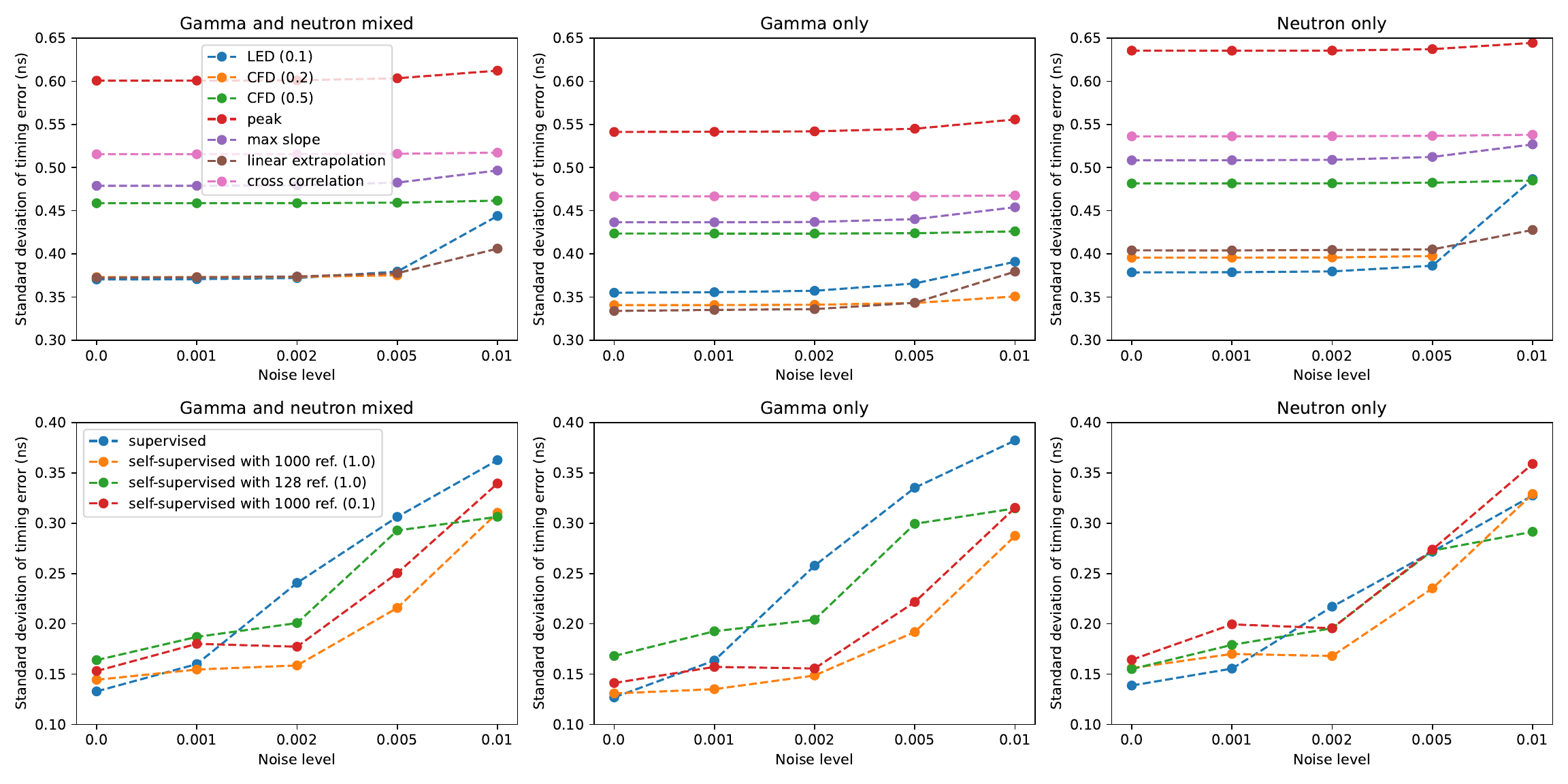}
	\caption{\label{fig:scint-sim} The timing results of the \emph{scintiPulses} simulation at different noise levels. We compare the conventional methods (top row) with the proposed neural regression models, for gamma/neutron-mixed events (left column), and gamma events (mid column) and neutron events (right column) separately. For methods with self-supervised learning, the numbers in the parentheses represent the coefficients for the weakly-supervised loss ($\lambda$ in equation (\ref{equ:hybrid-loss-func})). For LED (0.1), CFD (0.2) and CFD (0.5), the numbers in the parentheses represent the thresholds for these timing algorithms.}
\end{figure}

\begin{table}[htbp]
	\centering
	\begin{threeparttable}
		\caption{\label{tab:scint-sim-data} The timing results of the \emph{scintiPulses} simulation at different noise levels. We gather results from conventional methods and neural regression models trained by supervised learning and self-supervised learning with reference examples.}
		\begin{tabularx}{0.98\textwidth}{>{\raggedright\arraybackslash}Xp{0.35\textwidth}>{\centering\arraybackslash}X>{\centering\arraybackslash}X>{\centering\arraybackslash}X>{\centering\arraybackslash}X>{\centering\arraybackslash}X}
			\hline
			& & \multicolumn{5}{c}{Timing standard deviation (ns) at noise level} \\
			\cline{3-7}
			Category & Method & 0.0 & 0.001 & 0.002 & 0.005 & 0.01 \\
			\hline
			\multirow{11}{*}{Mixed} & supervised & \textbf{0.133} & 0.160 & 0.241 & 0.306 & 0.363 \\
			& self-supervised with 1000 ref. (1.0)\tnote{1} & 0.144 & \textbf{0.155} & \textbf{0.159} & \textbf{0.216} & 0.310 \\
			& self-supervised with 128 ref. (1.0)\tnote{1} & 0.164 & 0.187 & 0.201 & 0.293 & \textbf{0.306} \\
			& self-supervised with 1000 ref. (0.1)\tnote{1} & 0.153 & 0.180 & 0.177 & 0.250 & 0.339 \\
			& LED (0.1)\tnote{2} & 0.370 & 0.371 & 0.372 & 0.379 & 0.444 \\
			& CFD (0.2)\tnote{2} & 0.373 & 0.373 & 0.373 & 0.375 & --\tnote{3} \\
			& CFD (0.5)\tnote{2} & 0.459 & 0.459 & 0.459 & 0.459 & 0.462 \\
			& peak & 0.601 & 0.601 & 0.601 & 0.603 & 0.612 \\
			& max slope & 0.479 & 0.479 & 0.479 & 0.483 & 0.497 \\
			& linear extrapolation & 0.372 & 0.373 & 0.374 & 0.378 & 0.406 \\
			& cross correlation & 0.515 & 0.515 & 0.516 & 0.516 & 0.517 \\
			\hline
			\multirow{11}{*}{Gamma} & supervised & \textbf{0.127} & 0.164 & 0.258 & 0.335 & 0.382 \\
			& self-supervised with 1000 ref. (1.0)\tnote{1} & 0.131 & \textbf{0.135} & \textbf{0.149} & \textbf{0.192} & \textbf{0.287} \\
			& self-supervised with 128 ref. (1.0)\tnote{1} & 0.168 & 0.193 & 0.204 & 0.299 & 0.315 \\
			& self-supervised with 1000 ref. (0.1)\tnote{1} & 0.141 & 0.157 & 0.156 & 0.222 & 0.315 \\
			& LED (0.1)\tnote{2} & 0.355 & 0.356 & 0.357 & 0.366 & 0.391 \\
			& CFD (0.2)\tnote{2} & 0.341 & 0.341 & 0.341 & 0.343 & 0.351 \\
			& CFD (0.5)\tnote{2} & 0.424 & 0.424 & 0.424 & 0.424 & 0.426 \\
			& peak & 0.541 & 0.541 & 0.542 & 0.545 & 0.556 \\
			& max slope & 0.437 & 0.437 & 0.437 & 0.440 & 0.454 \\
			& linear extrapolation & 0.334 & 0.335 & 0.336 & 0.343 & 0.379 \\
			& cross correlation & 0.467 & 0.467 & 0.467 & 0.467 & 0.468 \\
			\hline
			\multirow{11}{*}{Neutron} & supervised & \textbf{0.139} & \textbf{0.156} & 0.217 & 0.272 & 0.327 \\
			& self-supervised with 1000 ref. (1.0)\tnote{1} & 0.156 & 0.170 & \textbf{0.168} & \textbf{0.235} & 0.329 \\
			& self-supervised with 128 ref. (1.0)\tnote{1} & 0.155 & 0.179 & 0.196 & 0.273 & \textbf{0.291} \\
			& self-supervised with 1000 ref. (0.1)\tnote{1} & 0.164 & 0.199 & 0.196 & 0.274 & 0.359 \\
			& LED (0.1)\tnote{2} & 0.379 & 0.379 & 0.380 & 0.386 & 0.487 \\
			& CFD (0.2)\tnote{2} & 0.396 & 0.396 & 0.396 & 0.398 & --\tnote{3} \\
			& CFD (0.5)\tnote{2} & 0.482 & 0.482 & 0.482 & 0.482 & 0.485 \\
			& peak & 0.636 & 0.635 & 0.636 & 0.637 & 0.644 \\
			& max slope & 0.508 & 0.509 & 0.509 & 0.512 & 0.527 \\
			& linear extrapolation & 0.404 & 0.404 & 0.405 & 0.405 & 0.428 \\
			& cross correlation & 0.536 & 0.536 & 0.536 & 0.537 & 0.538 \\
			\hline
		\end{tabularx}
		\begin{tablenotes}
			\footnotesize
			\item[1] The numbers in the parentheses represent the coefficients for the weakly-supervised loss ($\lambda$ in equation (\ref{equ:hybrid-loss-func})).
			\item[2] The numbers in the parentheses represent the thresholds for these timing algorithms.
			\item[3] We do not report the results of the digital CFD (0.2) at the noise level of 0.01 for Neutron and Mixed, because the timing algorithm fails to converge in these conditions.
		\end{tablenotes}
	\end{threeparttable}
\end{table}

In simulation, we totally generate 10000 examples, of which 5000 examples are gamma events and 5000 examples are neutron events. We randomly select 7000 examples as the training set and 1000 examples as the reference set, and the remaining 2000 examples are used for testing. It should be noted that, we have timing labels in the training set as well as in the reference set, but the labels in the training set are only used in supervised learning for comparison. In supervised learning, the training set and the reference set are gathered together for training. The waveform is sampled at 500 MSps. The length of the input time series is 256. A random range of 40 ns is applied to the starting point of each waveform. White Gaussian noise at different noise levels (noise standard deviation to average peak amplitude) is researched in this section. The loss function has been substantially reduced before we finish training.

The simulation results are shown in Fig.~\ref{fig:scint-sim} and Table~\ref{tab:scint-sim-data}. In total, 8 kinds of timing algorithms are being considered: supervised learning, self-supervised learning with weakly-supervised learning on reference examples (3 variants), LED, digital CFD (2 variants), peak finding, maximum slope, linear extrapolation, and cross correlation. For conventional timing algorithms, data preprocessing is performed before the algorithms are applied: the baseline is subtracted from the waveform, and the amplitude of the waveform is normalized to unity. For digital CFD, a delayed version of the pulse is compared to a decayed version to find the zero-crossing point as the predicted arrival time. For peak, maximum slope and cross correlation, an additional parabolic peak finding procedure is used to reach sub-period precision.

From the figure and table, it can be seen that, the methods based on machine learning display an advantage over conventional timing algorithms. In conventional algorithms, the LED, the digital CFD (0.2) and the linear extrapolation produce similar overall performance, whereas the digital CFD (0.5), the maximum slope, the cross correlation, and the peak finding give worse performance gradually. Besides, the timing performance of conventional algorithms is not sensitive to white Gaussian noise except for the case at the highest noise level, indicating that the limiting factor for these algorithms is not dominated by noise.

On the other hand, methods based on machine learning, reaching into lower regions of the standard deviation, are more sensitive to added noise. Besides, when noise level is low, supervised learning is better than the hybrid learning method; however, when noise level increases, the performance of supervised learning degrades faster than the hybrid learning method, and the latter performs better than the former. It is worth noting that, self-supervised learning with weakly-supervised learning on as few as 128 reference examples gives lower standard deviation than the hybrid learning method on 1000 reference examples in Neutron and Mixed conditions when the noise level is 0.01. This proves the importance of self-supervised learning in extremely high noise levels, which works as the main optimization target instead of a regularizer. Nevertheless, reducing the coefficient of the weakly-supervised loss does not have equivalent effects.

\section{Experimental results}
\label{sec:exp}

\begin{figure}[htb]
	\centering
	\includegraphics[width=0.98\textwidth]{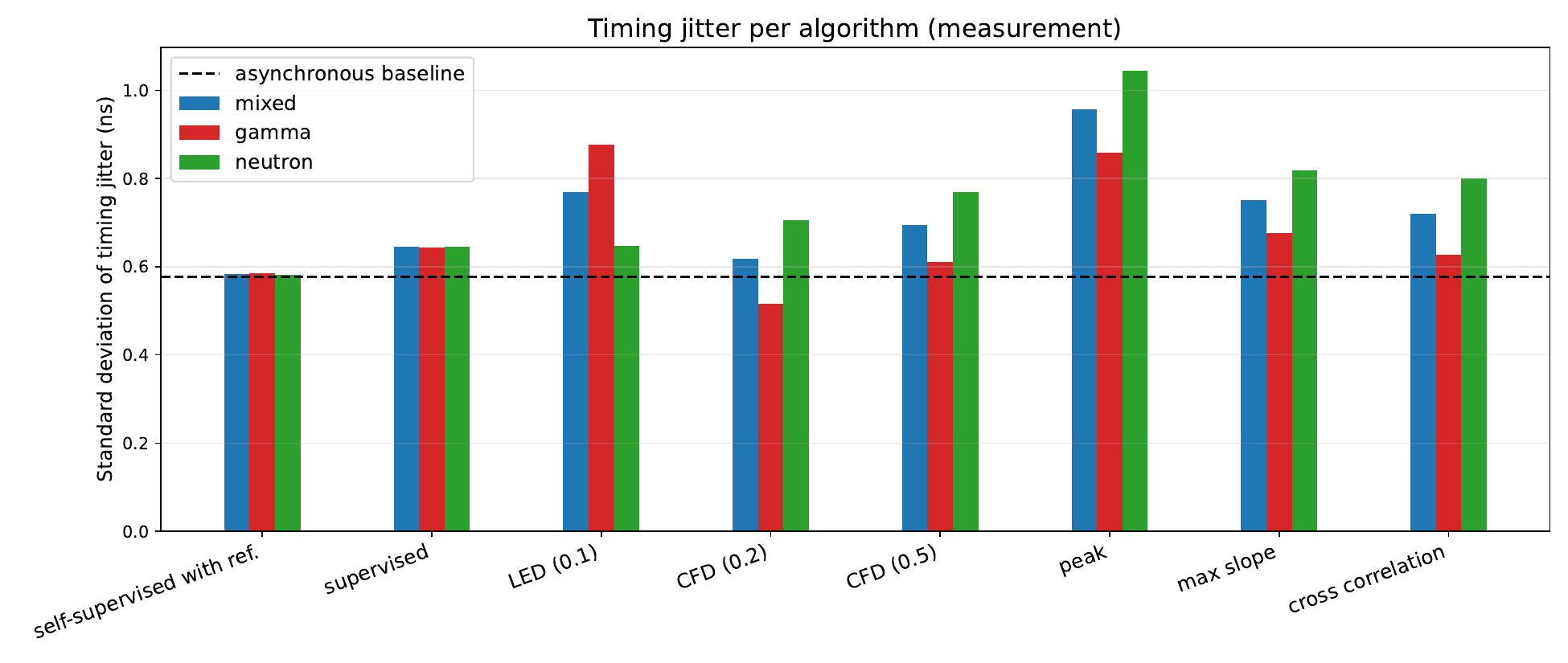}
	\caption{\label{fig:exp} The timing results of the ToF-labelled experimental data. We compare conventional methods and the proposed neural regression models on gamma/neutron-mixed events, and gamma events and neutron events separately. The default timing standard deviation caused by the asynchronous detection of events is indicated as a dashed line.}
\end{figure}

\begin{table}[htb]
	\centering
	\caption{\label{tab:exp} The timing results of the ToF-labelled experimental data. We gather results from conventional methods and neural regression models trained by supervised learning and self-supervised learning with reference examples. We do not include linear extrapolation in this table, because its performance is much poorer.}
	\begin{tabularx}{0.98\textwidth}{p{0.35\textwidth}>{\centering\arraybackslash}X>{\centering\arraybackslash}X>{\centering\arraybackslash}X}
		\hline
		& \multicolumn{3}{c}{Timing standard deviation (ns) per category} \\
		\cline{2-4}
		Method & Mixed & Gamma & Neutron \\
		\hline
		self-supervised with ref. & 0.583 & 0.585 & 0.582 \\
		supervised & 0.645 & 0.644 & 0.646 \\
		LED (0.1) & 0.770 & 0.877 & 0.646 \\
		CFD (0.2) & 0.617 & 0.515 & 0.705 \\
		CFD (0.5) & 0.695 & 0.611 & 0.769 \\
		peak & 0.957 & 0.859 & 1.045 \\
		max slope & 0.751 & 0.676 & 0.819 \\
		cross correlation & 0.720 & 0.627 & 0.801 \\
		\hline
	\end{tabularx}
\end{table}

We use the experimental dataset in a public repository \cite{PSDwithML} (also cited by Ref. \cite{Maedgen03102022}). Gamma and neutron events are generated by a ${}^{252}\text{Cf}$ source and discriminated by ToF measurements. The digitizer (CAEN DT5730) records the signals from the scintillator (ELJEN 309) at 500 MSps. Although the ToF information is utilized to prepare the dataset, the dataset does not include annotations of arrival time for recorded waveform signals. Hence, the timing jitter is used as the criterion to evaluate different timing algorithms.

The experimental dataset has 2197 gamma events and 2197 neutron events. We randomly select 3075 events (70\% of the dataset) as training examples and 439 events (10\% of the dataset) as reference examples. It should be noted that, for machine learning methods, the timing labels are acquired from the digital CFD (0.2) algorithm. This is not a perfect way to annotate examples, since timing uncertainties of the digital CFD are inherited, and the timing labels are inaccurate as a result. The length of the input time series is 256. A random range of 20 ns is applied to the starting point of each waveform. The loss function has been substantially reduced before we finish training.

The experimental results are shown in Fig.~\ref{fig:exp} and Table~\ref{tab:exp}. Since the source and the detector are asynchronous, the time when the particle reaches the detector can be any moment in a sampling period. This will introduce an inevitable standard deviation of the timing jitter originated from the variance of the uniform distribution:
\begin{equation}
	\sigma_{\mathrm{async}} = \sqrt{\frac{T_{\mathrm{period}}^2}{12}} = \frac{T_{\mathrm{period}}}{2\sqrt{3}} = \frac{2\ \mathrm{ns}}{2\sqrt{3}} = 0.577\ \mathrm{ns}
\end{equation}

\noindent where $T_{\mathrm{period}}$ is the sampling period of the digitizer. For an ideal timing algorithm, the standard deviation of the timing jitter should approximate the asynchronous baseline ($\sigma_{\mathrm{async}}$). It can be seen in the figure and the table that, although we do not use the accurate timing labels to train the hybrid learning method (which can be viewed as another type of weak supervision), the standard deviation of the timing jitter is very close to the asynchronous baseline, indicating the fitness of the proposed method. Among other methods, supervised learning gives slighter higher standard deviations than the asynchronous baseline; digital CFD (0.2) has similar performance as the hybrid learning method in the Mixed condition, but its standard deviation in the Gamma condition is actually lower than the asynchronous baseline, which is abnormal and can be viewed as a lack of variance instead of an advantage in precision; LED (0.1), digital CFD (0.5), peak finding, maximum slope, and cross correlation do not perform as well as the former three timing algorithms. To sum up, the experimental results demonstrate, by combining self-supervised learning and weakly-supervised learning, the proposed method is outstanding among researched timing algorithms, with the significantly alleviated cost in annotations of timing labels.

\section{Discussion}

Based on our observations during the study, the following points should be discussed:

\begin{itemize}
	\item In the hybrid loss function combining self-supervised learning and weakly-supervised learning, cautions should be exercised when gathering reference examples. To ensure the optimal performance, the reference examples and the training examples should come from an identical data distribution. Mismatches in the data distribution (also called ``concept drift'' in machine learning) will probably cause performance degradation.
	\item Throughout the study, we use a single neural network architecture as the predictor for pulse timing. In Ref. \cite{Ai_2023}, we have discussed the choices of different network architectures, including CNN, recurrent neural networks (RNN) like long short term memory (LSTM), and feedforward neural networks with dense connections. In general, CNN and LSTM are feasible choices, while CNN achieves better precision and LSTM achieves better linearity. If linearity is prioritized, LSTM is ready to replace the CNN architecture.
	\item Regardless of the hybrid learning method, deployment of the proposed timing predictor only involves a simple and compact CNN. When the input length of the CNN is 256, the size of weights is only 4.38 MB, which is suitable to be deployed on embedded edge devices, like microcontrollers with neural processing units (NPU) integrated. These devices are economical and widely available nowadays. When coupled with the front end of digitization, they can perform real-time and in situ feature extraction.
\end{itemize}

\section{Conclusion}

In this paper, we demonstrate that pulse timing, i.e. measuring the arrival time of electronic signals from radiation events, can be enhanced by self-supervised learning with neural regression models. Based on the theoretical analysis of the single-channel detector, we propose to combine self-supervised learning with weakly-supervised learning on a few reference examples and establish a new hybrid loss function. Proof-of-concept simulation investigates the time base drift problem when waveform examples have varied parameters, confirming the theoretical results. A thorough simulation and experimental study in a gamma/neutron-mixed radiation field illustrates the advantage of the proposed method, not only over conventional timing algorithms, but over supervised learning without self supervision. These results validate the outstanding performance of the hybrid learning method, which can be applied in nuclear facilities to implement real-time radiation monitoring and imaging. We sincerely hope the method proposed here can contribute to available radiation measurement techniques in the era of digital nuclear instrumentation.

\section*{CRediT authorship contribution statement}

\textbf{Pengcheng Ai:} Conceptualization, Software, Writing - Original Draft, Formal analysis, Investigation, Funding acquisition. \textbf{Xiangming Sun:} Writing - Review \& Editing, Supervision.

\section*{Declaration of competing interest}

The authors declare that they have no known competing financial interests or personal relationships that could have appeared to influence the work reported in this paper.

\section*{Acknowledgements}

This research is supported in part by the National Key Research and Development Program of China (under Grant No. 2024YFF0726201), in part by the National Natural Science Foundation of China (under Grant No. 12405228), and in part by the China Postdoctoral Science Foundation (under Grant No. 2023M731244).



\bibliographystyle{elsarticle-num} 
\bibliography{mybibfile}





\end{document}